\documentclass[,final]{aipproc}

\layoutstyle{6x9}

\newcommand{\fesc}{\mbox{$f_{\rm esc}$}}

\newcommand{\fion}[2]{\mbox{\rm [{#1}\thinspace{\footnotesize {#2}}]}} 

\newcommand{\Fhafuv}{\mbox{$F_{\rm H\alpha}/f_{\rm FUV}$}}

\newcommand{\gapeq}{\mbox{$~\stackrel{\scriptstyle >}{\scriptstyle \sim}~$}}
\newcommand{\HI}{\mbox {\sc Hi}}
\newcommand{\HII}{\mbox {\sc Hii}}
\newcommand{\HIPASS}{{\sc HiPASS}}
\newcommand{\Hline}[1]{\mbox{H{\footnotesize {#1}}}}
\newcommand{\Halpha}{\Hline{\mbox{$\alpha$}}}

\newcommand{\Mstar}{\mbox{${\cal M}_\star$}}
\newcommand{\Msun}{\mbox{${\cal M}_\odot$}}

\newcommand{\Mupp}{\mbox{${\cal M}_u$}}

\newcommand{\rxy}{\mbox{$r_{\rm xy}$}}
\newcommand{\SHI}{\mbox{$\Sigma_{\rm HI}$}}
\newcommand{\SSFR}{\mbox{$\Sigma_{\rm SFR}$}}
\newcommand{\SR}{\mbox{$\Sigma_R$}}
\newcommand{\tgas}{\mbox{$t_{\rm gas}$}}

\begin{document}

\title{The HI - Star Formation Connection: Open Questions}


\author{Gerhardt R. Meurer}{address={Department of Physics and
Astronomy, The Johns Hopkins University}}
\author{O.\ Ivy Wong}{address={School of Physics, University of
Melbourne, and Astronomy Department, Yale University}}
\author{Daniel J.\ Hanish}{address={Department of Physics and
Astronomy, The Johns Hopkins University}}

\begin{abstract}
We show data from the Survey of Ionization in Neutral Gas Galaxies
(SINGG) and Survey of Ultraviolet emission in Neutral Gas Galaxies
(SUNGG) which survey the star formation properties of \HI\ selected
galaxies as traced by \Halpha\ and ultraviolet emission,
respectively. The correlations found demonstrate a strong relationship
between the neutral ISM, young massive stars, and the evolved stellar
populations. For example the correlation between $R$ band surface
brightness and the \HI\ cycling time is tighter than the
Kennicutt-Schmidt Star Formation Law. Other scaling relations from
SINGG give strong direct confirmation of the downsizing scenario: low
mass galaxies are more gaseous and less evolved into stars than high
mass galaxies.  There are strong variations in the \Halpha\ to UV flux
ratios within and between galaxies.  The only plausible explanations for
this result are that either the escape fraction of ionizing photons or
the upper end of the IMF varies with galaxy mass.  We argue for the
latter interpretation, although either result has major implications for
astrophysics.  A detailed dissection of the massive star content in the
extended \HI\ disk of NGC~2915 provides a consistent picture of
continuing star formation with a truncated or steep IMF, while other
GALEX results indicate that star formation edges seen in \Halpha\ are
not always apparent in the UV.  These and other recent results settle
some old questions but open many new questions about star formation and
its relation to the ISM.
\end{abstract}

\maketitle

\section{Introduction}


Strong correlations between the star formation rate (SFR) of galaxies
and their \HI\ content have been known for some time.  For example
Kennicutt (1998a) showed that globally averaged star formation
intensity, \SSFR, in galaxies correlates more strongly with the \HI\
than with the CO surface density.  This is puzzling since stars form out
of the molecular not the neutral interstellar medium (ISM).

We are working on two surveys meant, in part, to examine the nature of
the \HI\ - star formation connection: the Survey of Ionization in
Neutral Gas Galaxies (SINGG), and the Survey of Ultraviolet emission in
Neutral Gas Galaxies (SUNGG).  These image nearby galaxies selected
blind to optical properties from the \HI\ Parkes All Sky Survey
(\HIPASS) in the light of two star formation tracers: \Halpha\ (SINGG)
and the far and near ultraviolet (FUV and NUV) continuum
(SUNGG). \Halpha\ emission traces the presence of ionizing O stars
having masses $\Mstar \gapeq\ 20\, \Msun$, while UV emission is
sensitive to both O and B stars having masses down to $\Mstar \gapeq\
3\, \Msun$.  Meurer et al.\ (2006) discuss the SINGG observations and
measurements. An initial description of the SUNGG survey can be found in
Wong (2007).  The full description of SUNGG is currently being written
by Wong et al.\ while the preliminary results presented here are being
written-up by Meurer et al.; both should be submitted for publication by
the middle of 2008.

Some of the open questions we aimed to address with these surveys
include What is the best form of the Star formation Law (SFL)?  Is it
constant?  Is the Initial Mass Function (IMF) universal?  What is the
heating source for the \HI\ dominated disks that are often seen to
extend well past the apparent optical extent of galaxies.

There have been many papers on the SFL.  Probably the most influential
have been the papers of Kennicutt and collaborators (Kennicutt, 1989;
Kennicutt 1998a, Martin \&\ Kennicutt 2001).  They show that the \SSFR\
has a power law dependence on the total ISM surface density $\SSFR
\propto \Sigma_g^N$ where $N \approx 1.4$, but only where the $\Sigma_g$
is large enough for the ISM disk to be self-gravitating.  Thus extended
\HI\ disks are thought to result if the ISM is not dense enough to form
stars.  The standard assumption in much of astronomy is that the IMF is
constant, which certainly seems to hold for stars clusters (Kroupa,
2001).  By using two star formation tracers we can test this assumption
and probe whether the same SFL that holds for O stars also works for B
stars.

The parameters derived from the SINGG and SUNGG data discussed here are
based on integrated fluxes measured from concentric elliptical
apertures.  In particular, \SSFR\ and the $R$ band surface brightness
\SR\ are measured within the half-light radius and corrected for
inclination.  The gas cycling time \tgas\ is derived from the ratio of
the \HI\ and \Halpha\ fluxes with a crude uniform correction for helium
and molecular gas.  Star formation rates are calculated using the
calibrations of Kennicutt (1998b) which adopt a Salpeter (1955) IMF over
the mass range of 0.1 to 100 \Msun.  The SINGG data are corrected for
dust absorption (and \fion{N}{II} contamination) using the relationships
of Helmboldt et al.\ (2004) and validated with FIR data as shown by
Meurer et al. (2006).  Dust absorption is removed from the UV fluxes
based on the FUV -- NUV colors (similar to Gil de Paz et al. 2007, for
example).

\section{Scaling relations and the Star Formation Law}

\begin{figure}
  \includegraphics[width=\textwidth]{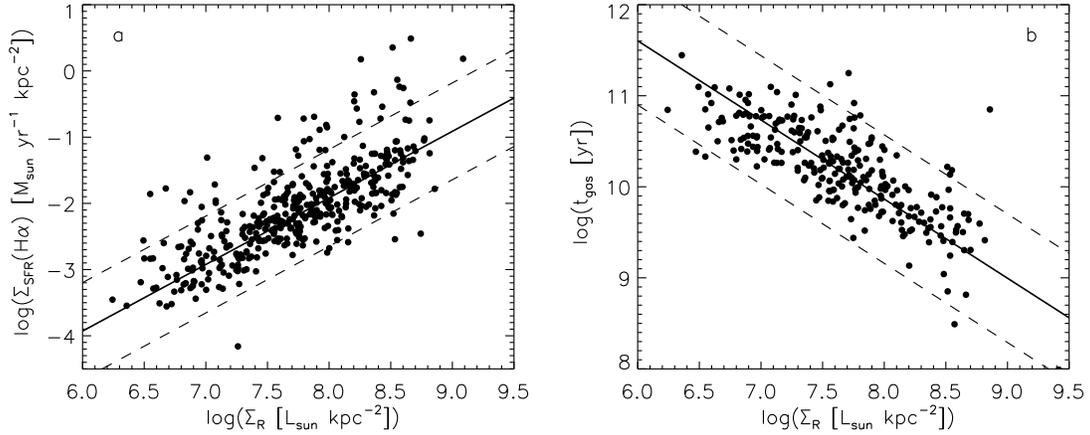} \caption{Correlations
  between (a) star formation intensity (derived from \Halpha), \SSFR,
  and the $R$ band surface brightness \SR, and (b) between the gas
  cycling time \tgas\ and \SR.  Solid lines show least squares fits with
  the outliers having residuals beyond $\pm 2.5\sigma$ iteratively
  clipped.  The clipping limits are shown as dashed lines.}
\label{f:srcorrs}
\end{figure}

Our surveys do not have the data necessary to fully recalibrate the SFL.
Instead we consider global properties and look for correlations, or
scaling relations. These amount to ``projections'' of the SFL.  As a
baseline for comparison, we fitted the disk averaged measurements
(ignoring central starbursts) from Kennicutt (1998a) to find a
correlation coefficient $r_{\rm xy} = 0.60$ and $N = 1.4$ for the
correlation between \SHI\ and \SSFR, with rms scatters in the residuals,
$\sigma$, of 0.40 dex in \SSFR.  The combined SINGG and \HIPASS\ data
allows us to recover the \SSFR\ versus \SHI\ correlation using a
``pseudo'' \HI\ surface density $\propto F_{\rm HI}/(\pi r_{90}^2)$,
where $ F_{\rm HI}$ is the \HI\ line flux, and $r_{90}$ is the radius
containing 90\%\ of the \Halpha\ flux.  Our correlation has $r_{\rm xy}
= 0.59$, $N = 1.66$, and $\sigma = 0.50$ dex in \SSFR.  Hence using a
pseudo-\SHI\ we recover a relationship very similar to the Kennicutt
SFL.

We find other strong correlations between parameters characterizing star
formation and the \HI\ and stellar content of the targets.
Figure~\ref{f:srcorrs}a shows a nearly linear correlation (slope $\alpha
= 1.08\pm 0.03$) between the \Halpha\ and $R$ band intensity.  This
relationship is {\em tighter\/} than the Kennicutt SFL and our \SSFR\
versus pseudo-\SHI\ relation discussed above.  Here $\rxy = 0.75$ and
the $\sigma = 0.29$ dex.  The average \Halpha\ equivalent width is
11\AA, while the $R$ band filter width is $\sim$1500\AA, hence \Halpha\
does not significantly contaminate the $R$ band and the strong
correlation is not spurious.  This implies that the older stellar
populations play at least as an important role in regulating star
formation as does the \HI\ content.  A similar result was reported by
Dopita \&\ Ryder (1994).  Note the many outliers above the correlation,
these correspond to starburst systems.  An even tighter (anti-)
correlation, with fewer outliers, exists between \tgas\ and \SR\ as
shown in Fig.~\ref{f:srcorrs}b.  Here $\rxy = -0.77$ and $\sigma = 0.28$
dex.  Using this correlation ``in reverse'' one can estimate the \HI\
content of galaxies to better than a factor of 2 from $R$ and \Halpha\
measurements.

\begin{figure}
  \includegraphics[width=\textwidth]{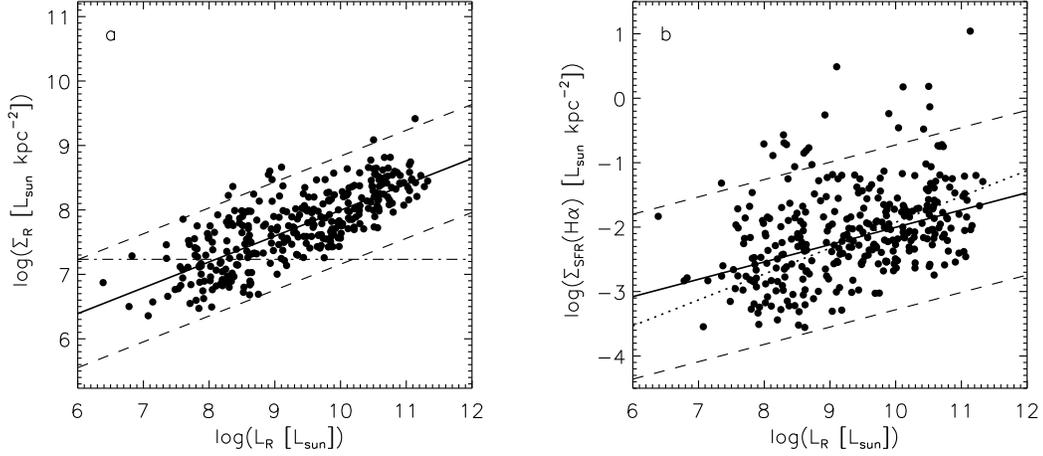} \caption{The
  correlations between $R$ band luminosity and (a) $R$ band surface
  brightness and (b) star formation intensity.  Symbols and line styles
  are the same as in Fig.~\ref{f:srcorrs}.  The dot-dashed
  line in panel (a) shows the approximate surface brightness limit of
  the SDSS spectroscopic survey, while the dotted line in panel (b) has
  the same slope as the fit shown in panel (a). }
\label{f:lumsb}
\end{figure}

Figure~\ref{f:lumsb} shows the luminosity - surface brightness
relationship of our sample.  In panel (a) the $R$ band relationship is
shown.  Here $\rxy = 0.76$ and $\alpha = 0.40\pm 0.02$, $0.53\pm0.02$
for ordinary least squares fit of Y on X and bisector methods,
respectively.  Fitting of the former type is displayed here because it
works better when the correlation is weak as in panel (b).  The
bisector results agree very well with results found using over $10^5$
galaxies from the Sloan Digital Sky Survey by Kauffmann et al. (2003).
In terms of stellar mass $\Mstar$ and mass density $\Sigma_\star$ they
find $\Sigma_\star \propto \Mstar^{0.54\pm 0.03}$, with $\sigma \approx
0.2$ dex, considerably tighter than our relationship which has $\sigma =
0.36$ dex.  However, we can trace the correlation to fainter intensities
using the SINGG data because our results do not require optical
spectroscopy (see the dot-dashed line in Fig.~\ref{f:lumsb}).

In \Halpha, the luminosity - surface brightness correlation is weaker
and shallower than in the $R$ band as shown in Fig.~\ref{f:lumsb}b.
Here $\rxy = 0.38$, $\alpha = 0.27 \pm 0.03$ and $\sigma = 0.51$ dex,
again with high surface brightness starburst outliers apparent.  The
relative shallowness of the correlation when using \Halpha\ is
consistent with the expectations of the ``down-sizing'' scenario (Cowie 1996);
relative to the stars already in place the star formation activity is
stronger for lower luminosity systems.  Stronger evidence of downsizing
in the SINGG sample was presented by Hanish et al. (2006) who show that
\HI\ contributes a larger fraction of the dynamical mass than stars for
low mass galaxies, while the situation is reversed for high mass
galaxies.  Low mass galaxies are less evolved because they have
converted less of their ISM into stars than high mass galaxies.

\section{The H$\alpha$/FUV flux ratio and the Initial Mass Function}

\begin{figure}
  \includegraphics[width=\textwidth]{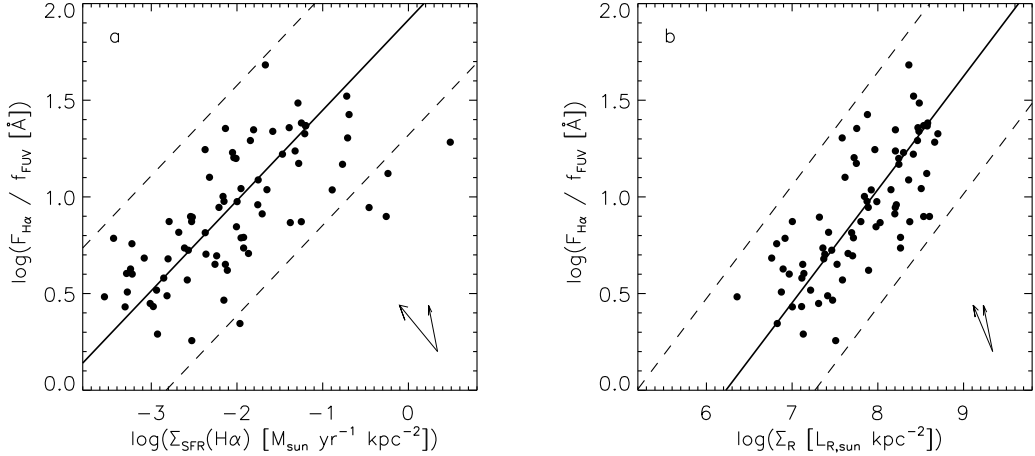} \caption{The
  correlations between the \Halpha\ to FUV flux ratio \Fhafuv\ and (a)
  star formation intensity, as derived from \Halpha\, and (b) $R$ band
  surface brightness.  Symbols and line styles are the same as in
  Fig.~\ref{f:srcorrs}.  The vectors in the bottom right show the
  effects of dust reddening for two different models. The more vertical
  of the vectors is the Galactic dust extinction model (Cardelli et al.\ 1989); the
  other is the Calzetti (2000) Starburst attenuation law.  The
  length of both vectors in $\log(\Fhafuv)$ is set to 0.24 dex which
  corresponds to an attenuation of the $V$ band stellar continuum of
  $A_V = 0.33$, 0.68 mag for Galactic and Starburst reddening
  respectively.  }
\label{f:hafuvsb}
\end{figure}

Figure~\ref{f:hafuvsb} shows strong correlations between the ratio of
\Halpha\ line flux to FUV continuum flux density, \Fhafuv, and (a)
\SSFR, and (b) \SR.  In panel a (b) the correlation coefficient
\rxy, slope $\alpha$, and dispersion of residuals $\sigma$ are
0.67, 0.47, 0.24 (0.74, 0.59, 0.24) respectively.  What drives these
strong correlations?  The \Fhafuv\ ratio is very sensitive to the O to B
star ratio.  The O to B star ratio is in turn affected by the parameters
specifying the upper end of the IMF, and the star formation history.
Other parameters that affect \Fhafuv\ are the dust extinction, and the
escape fraction \fesc\ of ionizing radiation.  We posit that systematic
variation of the IMF parameters are the most likely cause of the
correlations seen in Fig.~\ref{f:hafuvsb}.

If there are residual errors in our dust absorption correction then the
effect will be to move data along trajectories roughly perpendicular to
the observed correlations as shown by the reddening vectors in
Fig.~\ref{f:hafuvsb}.  The observed correlation can not be created by
stretching out uncorrelated data with a spurious extinction correction.
Addition of a starburst can cause a short term increase in both \Fhafuv\
and \SSFR; likewise sharply truncated star formation can cause a
decrease in both quantities.  However, for star formation history to be
behind the observed correlations requires almost as much change in \SR\
as \SSFR\ which is not possible with plausible population models.  The
strongest argument against the star formation history scenario is that
\Fhafuv\ also correlates with other global quantities such $L_R$ (as may
be inferred from Fig.~\ref{f:lumsb}a) and dynamical mass.  Galaxies with
low \Fhafuv\ tend to be low surface brightness dwarf galaxies, while
galaxies with high \Fhafuv\ tend to be high luminosity massive spirals.
Such a dramatic range of properties can not be acquired by short term
changes in the SFR.

Having \fesc\ inversely correlated with galaxy mass could cause the
observed correlations.  \fesc\ has only been well measured in high
surface brightness galaxies.  So we can not yet rule out a variable
\fesc\ as the cause of the observed correlations.  This would require
that dwarf galaxies have an ISM that is more porous to ionizing photons
than in more massive galaxies.  Dwarf galaxies with porous \HI\
distributions have been observed (e.g.\ Puche et al. 1992).  However
dwarf galaxies typically contain a higher fraction of their mass in the
ISM (Hanish et al. 2006), while the lower mass densities of their disks
suggests that that the ISM distribution should be ``puffier''.  These
factors should make it harder for ionizing photons to escape, not
easier.  Indeed, Oey et al. (2007) argue that \fesc\ increases with
surface brightness.  

The observed range in \Fhafuv\ can be accounted for by plausible
adjustments to the parameters affecting the upper end of the IMF. Using
Starburst99 population synthesis models (Leitherer et al.\ 1999,
V{\'a}zquez \&\ Leitherer 2005) having constant SFR and solar
metallicity we find that the range of \Fhafuv\ can be modeled with an
IMF having a Salpeter slope, $\gamma = -2.35$, if the upper mass
limit \Mupp\ varies between 30 and 120 \Msun, or for a fixed $\Mupp =
80\, \Msun$ if $\gamma$ varies between --1.1 and --3.5.  What is
required is that one or both of these properties varies systematically
with global galaxy properties.

\section{Extended HI disks}


In separate work with the Advanced Camera for Survey (ACS) team, we
looked for young stars in the extremely extended outer \HI\ disk of
NGC~2915 (Meurer et al. 1996) in two ways.  First, we imaged the galaxy
in \Halpha\ with the Anglo-Australian Telescope and discovered three
faint \HII\ regions beyond the Holmberg isophote.  Each of these could
be ionized by a single late O or early B star.  Second, we used ACS on
the Hubble Space Telescope to take a deeper look for stellar populations
in the outer disk, selecting a field that contained one of the outer
\HII\ regions found from the ground.  We found a pervasive distribution
of old stars including three globular clusters (Meurer et al. 2003) as
well as a smattering of blue stars and an open cluster at the position
of the \HII\ region. The detection limit of the data corresponds to a
mass limit of about 7 \Msun\ on the main sequence, hence the blue stars
are primarily B stars.  The distribution of blue stars is flat in a
radial sense, but clumpy and highly correlated with the \HI.  We found a
total of 430 blue stars, which must be predominantly on the main
sequence.  For a constant SFR population having a Salpeter IMF with
$\Mupp = 100$ \Msun\ one expects an equilibrium B/O star number ratio of
22.  The AAT \Halpha\ observations allow 1-4 O or late B stars.  Hence,
the B/O ratio is too high for this IMF by a factor of 5-20.  The \HI\
morphology of NGC~2915 is very regular, while the orbital time at the
radii probed by the HST observations is $\sim$ 200 Myr, longer than the
lifetimes of the stars observed.  The high B/O ratio can not be caused
by a truncated star formation history - you can't turn off the galaxy
quick enough.  The brightest blue stars are at the position of the \HII\
region in the field.  There is no evidence for ``naked'' O stars in the
field.  Hence the best explanation for the high B/O ratio in this field
is a steep $\gamma$ or low \Mupp.

Other examples of O star deficient outer disks have also been reported
recently in the literature.  A spectacular example is M83 which was used
as the primary example of a galaxy with a star formation edge deduced
from \Halpha\ observations by Martin \&\ Kennicutt (2001).  However
Thilker et al.\ (2005) used GALEX observations to demonstrate that the
UV emission extends much further than the \Halpha\ light with no sign of
a star formation-edge in the UV radial profiles.  

\section{Conclusions}

Recent work by our teams as well as others has cleared up several open
questions.  The star formation, \HI\ and stellar light properties of
galaxies are tightly correlated as shown by global scaling relations,
indicating that an improved form of the star formation law is within
reach.  This must have the star formation rate dependent on the stellar
as well as ISM mass density, perhaps similar to the form suggested by
Dopita \&\ Ryder (1994).  It makes sense that the stellar mass density
should contribute to regulating star formation since in most galaxies
stars are the major contributor to the galactic potential of the
optically bright portion of galaxies, and hence are key to setting the
hydrostatic pressure of galactic disks.  It appears that extended \HI\
disks are not empty of stars but have sparse populations of B stars that
heat the disk.  It is also becoming clear that FUV and \Halpha\
properties are different between galaxies and even have different
distributions within galaxies.  This probably indicates that the B/O
ratio varies within and between galaxies, and the most likely
explanation for that is the upper end of the IMF is not universal.  
The only alternative is that the escape fraction of ionizing photons is
much larger at the low surface brightness (low mass) end of the star
forming galaxy sequence.  While this is not ruled out by observations, it
is contrary to naieve expectations.  

Whatever the cause of the systematic \Fhafuv\ variations there are major
implications and many new questions to resolve.  What is \fesc\ in low
surface brightness galaxies?  If the IMF varies, which parameters vary
and what drives the variations?  What is the best way to measure the SFR of
galaxies?  Other open questions relate more directly to the
\HI\ - star formation connection.  On the most basic level what is the
nature of the connection?  Is \HI\ a tracer for the ISM that fuels the
star formation, or does it represent the byproduct of the young stellar
populations photo-dissociating the molecular ISM they formed out of
(e.g.\ Tilanus \&\ Allen 1993)?  Finally, what form of the SFL best
explains the inter-relationship between star formation, the existing
stars, and the \HI\ content of galaxies?

\begin{theacknowledgments}

Combined the SINGG, SUNGG and ACS Instrument Definition Teams includes
69 members, all of which contributed to making these large projects
function.  Unfortunately there is insufficient space to acknowledge them
fully here.  This work was supported by NASA LTSA grant NAG5-13083, NASA
Galex Guest Investigator grant GALEXGI04-0105-0009, and NASA grant
NAG5-7697.
\end{theacknowledgments}





\end{document}